\documentclass{article}
\usepackage[dvips]{graphicx}  
\usepackage{color}

\usepackage[varg]{txfonts}
\usepackage{concmath}

\begin{document}

\newcommand{\rum}{\rule{0.5pt}{0pt}}
\newcommand{\rub}{\rule{1pt}{0pt}}
\newcommand{\rim}{\rule{0.3pt}{0pt}}
\newcommand{\numtimes}{\mbox{\raisebox{1.5pt}{${\scriptscriptstyle \rum\times}$}}}
\newcommand{\numtimess}{\mbox{\raisebox{1.0pt}{${\scriptscriptstyle \rum\times}$}}}
\newcommand{\Boldsq}{\vbox{\hrule height 0.7pt
\hbox{\vrule width 0.7pt \phantom{\footnotesize T}%
\vrule width 0.7pt}\hrule height 0.7pt}}
\newcommand{\two}{$\raise.5ex\hbox{$\scriptstyle 1$}\kern-.1em/
\kern-.15em\lower.25ex\hbox{$\scriptstyle 2$}$}

\renewcommand{\refname}{References}
\renewcommand{\tablename}{\small Table}
\renewcommand{\figurename}{\small Fig.}
\renewcommand{\contentsname}{Contents}

\twocolumn[%
\begin{center}
{\Large\bf 
Dynamical 3-Space Gravitational Waves: Reverberation Effect \rule{0pt}{13pt}}\par

\bigskip
Reginald T. Cahill  and Samuel T. Deane\\ 
\vspace{1mm}Progress in Physics   {\bf 2}, 9-11, 2013\\
{\small\it  School of Chemical and Physical  Sciences, Flinders University,
Adelaide 5001, Australia\rule{0pt}{15pt}}\\
\raisebox{+1pt}{\footnotesize E-mail: Reg.Cahill@flinders.edu.au}\par

\bigskip

\bigskip

{\small\parbox{11cm}{%
Gravity theory missed a key dynamical process that became apparent only when expressed in terms of a velocity field, instead of the Newtonian gravitational acceleration field.  This dynamical process involves an additional  self-interaction of the dynamical 3-space, and experimental data reveals that its strength is set by the fine structure constant, implying a fundamental link between gravity and quantum theory.  The dynamical 3-space has been directly detected in numerous light-speed anisotropy experiments.  Quantum matter has been shown to exhibit an acceleration caused by the time-dependence and inhomogeneity of the 3-space flow, giving the first derivation of gravity from a deeper theory, as a quantum wave refraction effect.   EM radiation is also refracted in a similar manner. The anisotropy experiments have all shown 3-space wave/turbulence effects, with the latest revealing the fractal structure of 3-space. Here we report the prediction of a new effect, namely a reverberation effect, when the gravitational waves  propagate in the 3-space inflow of a large mass. This effect arises from the non-linear dynamics of 3-space. These reverberations could offer an explanation for the Shnoll effect, in which cosmological factors influence stochastic processes, such as radioactive decay rates.
\rule[0pt]{0pt}{0pt}}}\medskip
\end{center}]{%

\setcounter{section}{0}
\setcounter{equation}{0}
\setcounter{figure}{0}
\setcounter{table}{0}

\markboth{Cahill R.T. and  Deane S.T.   Dynamical 3-Space Gravitational Waves:  Reverberation Effects }{\thepage}
\markright{Cahill R.T.  and   Deane S.T.   Dynamical 3-Space Gravitational Waves: Reverberation Effects }

\section{Introduction}
Newton's inverse square law of gravity, when expressed in terms of an acceleration field ${\bf g}({\bf r},t)$, has the differential form:
\begin{equation}
\nabla.{\bf g}=-4\pi G\rho,  \mbox{\ \ \  } \nabla \times {\bf g} = {\bf 0} 
\label{eqn:Newton}\end{equation}
where G is the gravitational constant and $\rho$ is the real matter density. The $g$ field was believed to exist within an actual  Euclidean  space. It has become increasingly evident through the observation of spiral galaxies, the expanding universe and gravitational anomalies, that Newton's inverse square law is an incomplete theory of gravity.  However a unique generalisation of  (\ref{eqn:Newton})  has lead to a resolution of these anomalies, by writing the acceleration field ${\bf g}({\bf r},t)$ in terms of the Euler acceleration of a velocity field ${\bf v}({\bf r},t)$ \cite{CahillBook,Review}:
\begin{equation}
{\bf g}=\frac{\partial{\bf v}}{\partial t}+({\bf v}.\nabla){\bf v},
\label{eqn:acceln}\end{equation}
\begin{equation}
\nabla.\left(\frac{\partial {\bf v} }{\partial t}+ ({\bf v}.{\bf \nabla}){\bf v}\right)
=-4\pi G\rho,  \mbox{\ \ \  } \nabla \times {\bf v} = {\bf 0}. 
\label{eqn:Newtonv} \end{equation}
This approach utilises the the well known Galilean covariant Euler acceleration for a fluidic flow of the substratum with velocity  ${\bf v}({\bf r},t)$.  The velocity field is defined relative to an observer. The time dependent nature of the flow means that Newtonian gravity, within  this flow formalism, can support wave phenomena.  But a unique term can be added to (\ref{eqn:Newtonv})  that generalises the flow equation, but also  preserves the Keplerian nature of the planetary motions that underlie Newton's gravity formalisation:
 \begin{eqnarray}
\nabla.\left(\frac{\partial {\bf v} }{\partial t}+ ({\bf v}.{\bf \nabla}){\bf v}\right)+
\frac{\alpha}{8}\left((tr D)^2 -tr(D^2)\right)=-4\pi G\rho, \nonumber\\
 \nabla\times {\bf v}={\bf 0},  \mbox{\  \  \   }
 D_{ij}=\frac{1}{2}\left(\frac{\partial v_i}{\partial x_j}+
\frac{\partial v_j}{\partial x_i}\right).
\label{eqn:3spacedynamics}
\end{eqnarray}
  Analysis of Bore Hole $g$ anomaly data  revealed that $\alpha$ is the fine structure constant \cite{CahillBook}. The additional dynamics explains the  ``dark matter" effects, and so may be referred to as the dark matter term:
\begin{equation}
\rho_{DM}({\bf r},t)=\frac{\alpha}{32\pi G}\left((tr D)^2 - tr(D^2)\right)
\end{equation}
whereby 
\begin{equation}
\nabla\cdot {\bf g}=\nabla\cdot\left(\frac{\partial \bf v}{\partial t}+\nabla\left(\frac{{\bf v}^2}{2}\right)\right)=-4\pi G\left(\rho_M +\rho_{DM}\right)
\end{equation}
Dynamical 3-Space is unlike the dualistic space and aether theories of the past, as herein only space exists, and there is no aether.  This dynamical and structured space provides an observable and observed local frame of reference. The flow of  the dynamical 3-space has been detected many times dating back to the Michelson and Morley  1887 experiment,  which has always,  until 2002,  been mistakenly reported as a null experiment.   Wave effects, essentially  gravitational waves, are apparent in the data from various anisotropy experiments\footnote{Vacuum mode Michelson interferometers have zero sensitivity to these waves. So such devices have a fundamental design flaw when intended to detect such waves.}.   A large part of understanding gravitational waves lies in how they originate, and also in understanding how they propagate. This work herein investigated the propagation of these gravitational waves within the background in-flow of a large mass, such as the earth or the sun.  In doing so it was discovered that the dynamics of the propagation resulted in a reverberation effect, caused by the non-linear nature of the flow dynamics, apparent in (\ref{eqn:Newtonv}) and (\ref{eqn:3spacedynamics}).

\begin{figure}[h]
\centering
\includegraphics[width=0.5\textwidth]{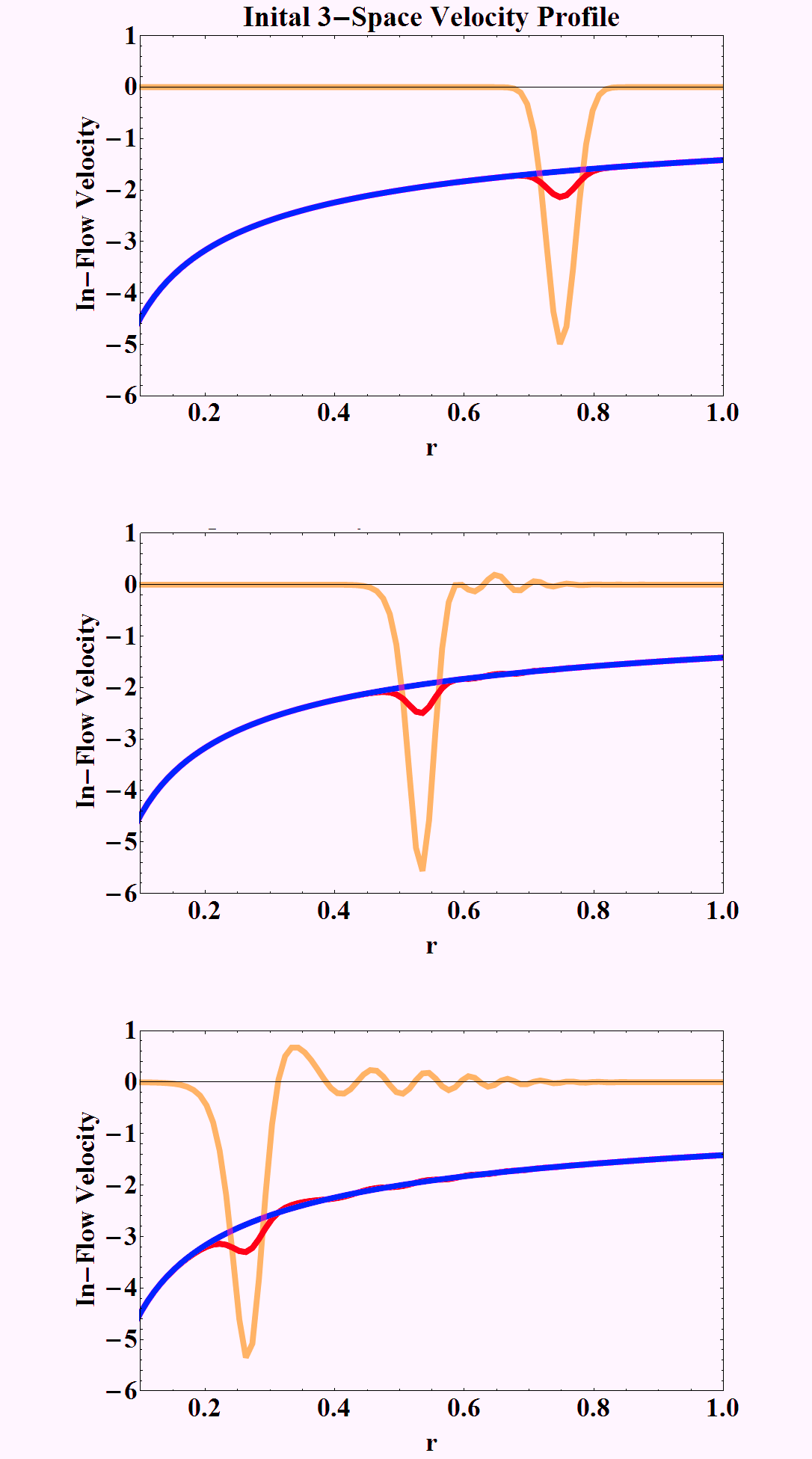}
\caption{Inflowing 3-space perturbation $w(r,t)$  (red) and un-perturbed inflow $v(r)$ (blue) velocity profiles outside a mass,  with the  waveform $w(r,t)$ also shown magnified (yellow), at later times.}\label{fig:reverberations}
\end{figure}

\section{3-Space Flow Dynamics}
First we establish the in-flow of space into a spherical mass, assuming for simplicity that the mass is asymptotically at rest, which means that the in-flow has spherical symmetry.  In the case of the earth we know that the earth has a large velocity wrt  to the local 3-space frame of reference, some 486km/s in the direction RA = 4.3$^{h}$, Dec = -75$^\circ$, \cite{CahillNASA}.   Here we restrict the analysis to the case of a spherically symmetric inflow into a spherical mass, with density $\rho(r)$ and total mass $M$.  Then   (\ref{eqn:3spacedynamics}) becomes  ($v^\prime \equiv \partial v(r,t)/\partial r$)
\begin{equation}
\frac{\partial v^\prime}{\partial t}+
v v^{\prime\prime}+\frac{2vv^\prime}{r}+(v^\prime)^2+\frac{\alpha}{2}\left(\frac{v^2}{2r^2}+\frac{vv^\prime}{r}\right)=-4\pi G\rho 
\label{eqn:1D}\end{equation}
which for a static flow  has the exact solution 
\begin{eqnarray}
v(r)^2=
& \displaystyle {\frac{2\beta}{r^\frac{\alpha}{2}} + \frac{2G}{(1-\frac{\alpha}{2})r} \int_0^r 4 \pi s^2 \rho(s) ds}    \nonumber  \\
  +& \displaystyle {\frac{2G}{(1-\frac{\alpha}{2})r^\frac{\alpha}{2}} \int_r^\infty 4 \pi s^{1+\frac{\alpha}{2}} \rho(s) ds} ,\label{eqn:RawGravitySol1}
\end{eqnarray}
Here $M$ is the total matter mass, and $\beta$ is a free parameter. The term $2\beta/r^{\alpha/2}$ describes an inflow singularity or ``black hole" with arbitrary strength. This is unrelated to the putative black holes of General Relativity. This corresponds to a primordial black hole.  As well the last term in  (\ref{eqn:RawGravitySol1})
also has a $1/r^{\alpha/2}$ inflow-singularity, but whose strength is mandated by the matter density, and is absent when $\rho(r)=0$ everywhere. This is a minimal ``black hole", and is present in all matter systems.  The  $2\beta/r^{\alpha/2}$ term will produce a novel long range gravitational acceleration $g=\alpha\beta/2r^{1+\alpha/2}$, as observed in spiral galaxies.  For the region outside the sun Keplerian orbits are known to well describe the motion of the planets within the solar system, apart from some small corrections, such as the Precession of the Perihelion of Mercury, which follow from relativistic effects in the more general form of (\ref{eqn:acceln}), \cite{CahillBook}.   The case $\beta=0$ applies to  the sun and earth, having only   induced  ``Minimal Attractor" back holes.  These minimal black holes contribute to the external $g(r)=GM^*/r^2$ gravitational acceleration, through an effective mass
\begin{equation}
M^* \approx M+\frac{\alpha}{2} M
\label{eqn:BHmass}\end{equation} 

Outside of a spherical mass, with only an induced black hole, (\ref{eqn:RawGravitySol1}) has a solution $v \propto 1/\sqrt{r}$, and then $\rho_{DM} = 0 $ outside of the sphere, which explains why the $\alpha-$term in (\ref{eqn:3spacedynamics}) went undiscovered until 2005.

\section{Gravitational Wave Reverberations}
We now demonstrate that  gravitational waves incoming on,  say, a star or planet develop reverberations, in which the wave generates following copies of itself.  For numerical accuracy in solving 
for time dependent effects in (\ref{eqn:3spacedynamics}), we assume a spherically symmetric incoming wave, which is clearly unrealistic, and so find numerical solutions to (\ref{eqn:1D}), by using the ansatz $v(r,t)=v(r)+w(r,t)$, where $v(r) \sim -1/\sqrt{r}$ is the static in-flow from  (\ref{eqn:RawGravitySol1}), applicable outside of the star/planet, and so ignoring the galactic background flow, and where $w(r,t)$ is the wave effect, with the initial wave $w(r,0)$ having the form of a pulse, as shown in fig.\ref{fig:reverberations}, where  the time evolution of $w(r,t)$ is  also shown.   We see that the initial pulse develops following copies of itself.  This is a direct consequence of the non-linearity of  (\ref{eqn:3spacedynamics}), or even (\ref{eqn:Newtonv}).

These reverberations are expected to  be detectable in EM speed anisotropy experiments.  However because the 3-space is fractal, as illustrated in fig.\ref{fig:Space},  \cite{CahillWaves},
the reverberations are expected to be complex.  As well all systems would generate reverberations, from  planets, moons, sun and the galaxy.  The timescale for such reverberations would vary considerably.  As well as being directly observable in EM anisotropy and gravitational wave detectors, these reverberations would affect, for example, nuclear decay rates, as the magnitude of the 3-space fractal structure is modulated by the reverberations, and this fractal structure will stimulate nuclear processes.  Patterns in the decay rates  of nuclei have been observed  by Shnoll {\it et al.}, with periodicities over many time scales \cite{Shnoll}.

The 3-space is detectable because the speed of EM waves, in vacuum is $c \approx 300,000$km/s wrt that space itself,  whereas an observer, in general, will observe anisotropy when the observer is in motion wrt the space.  This effect has been repeatedly observed for over 120 years.  The anisotropy   detections have always revealed wave/turbulence effects, including the original Michelson-Morley experiment. These wave effects are known as  ``gravitational waves", although a more appropriate description would be ``space waves''.  In the limit $\alpha\rightarrow 0$  (\ref{eqn:3spacedynamics}), and hence also (\ref{eqn:Newtonv}), still has space wave effects, but these generate gravitational wave effects, namely fluctuations in the matter acceleration field $g(r,t)$, only when  $\alpha \neq 0$. So the $\alpha$-dynamical term is not only responsible for the earth bore hole $g$ anomaly, and  for the so-called ``dark matter" effects in spiral galaxies, but can also result in forces acting on matter resulting from the space wave phenomena, and will be large when significant wave effects occur, with large wave effects being essentially a galactic effect.

 \begin{figure}
\hspace{1mm}\includegraphics[scale=0.35]{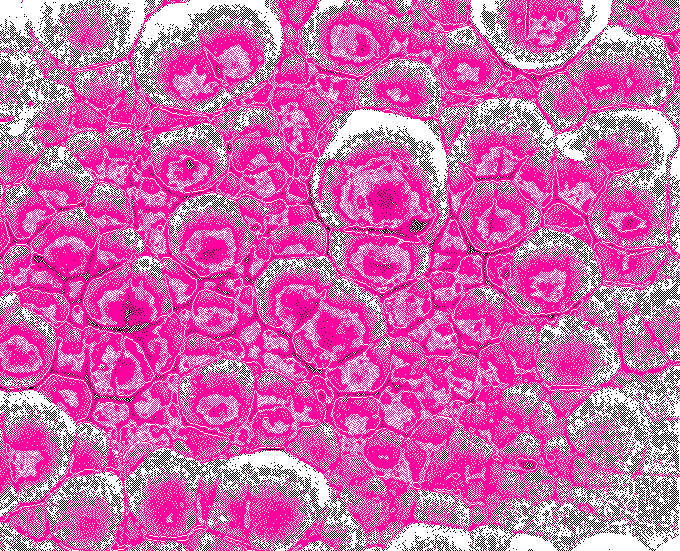}
	\caption{\small {Representation of the wave data   revealing the fractal textured structure of the 3-space, with cells of space having slightly different velocities, and continually changing, and moving wrt the earth with a speed of $\sim$500km/s, from \cite{CahillWaves} .  }}
\label{fig:Space}\end{figure}

\section{Acknowledgements}
This report is from the Flinders University Gravitational Wave Detector Project,  funded by an Australian Research Council Discovery Grant: {\it Development and Study of a New Theory of Gravity}.  
Special thanks to  Dr David Brotherton-Ratcliffe and Professor Igor Bray for ongoing support.

\end{document}